\documentclass{egpubl}
\usepackage{eurovis2026}

 \SpecialIssueSubmission    

\CGFccby

\usepackage[T1]{fontenc}
\usepackage{dfadobe}  

\usepackage{cite}  
\BibtexOrBiblatex
\electronicVersion
\PrintedOrElectronic
\ifpdf \usepackage[pdftex]{graphicx} \pdfcompresslevel=9
\else \usepackage[dvips]{graphicx} \fi

\usepackage{egweblnk}

\usepackage[dvipsnames]{xcolor}

\newcommand{\rev}[1]{{\textcolor{black}{\normalsize {#1}}}}
\usepackage{egweblnk}
\usepackage{subcaption}
\usepackage{graphicx}

\newcommand{\wib}[1]{{#1{orkshop-in-a-box}}}
\newcommand{\akbaba}{Derya}
\newcommand{\svensson}{Camilla}
\newcommand{\torelli}{Claudia}
\newcommand{\callmeryd}{Martin}
\newcommand{\meyer}{Miriah}
\newcommand{\organization}{Stadsmission}
\newcommand{\sweden}{Sweden}
\newcommand{\swedish}{Swedish}
\newcommand{\pidh}{Power in da Hood}
\newcommand{\visc}{Visualiseringscenter C}

\definecolor{boxcolor}{HTML}{DEB887}
\definecolor{boxbordercolor}{HTML}{8B4513}
\newcommand{\wibbox}[1]{
    \fcolorbox{boxbordercolor}{boxcolor}{
        \begin{minipage}{0.87\linewidth}
            #1
        \end{minipage}
    }
    \\
}

\definecolor{boxpcolor}{HTML}{E49BA6}
\definecolor{boxpbordercolor}{HTML}{92487A}
\newcommand{\proceduresbox}[1]{
    \fcolorbox{boxpbordercolor}{boxpcolor}{
        \begin{minipage}{0.87\linewidth}
            #1
        \end{minipage}
    }
    \\
}

\newcommand{\wibBOX}[2]{{
    \par
    \noindent
    \begin{minipage}{0.87\linewidth} 
    
        \fcolorbox{boxbordercolor}{boxcolor}{\parbox{\linewidth}{#1}}
        \vspace*{-0.4pt} 
        \fcolorbox{boxpbordercolor}{boxpcolor}{\parbox{\linewidth}{#2}}
        
    \end{minipage}
    \par 
}}

\usepackage{titlesec}
\titlespacing*{\section}{0pt}{1em}{-0.5em}
\titlespacing*{\subsection}{0pt}{1em}{-0.5em}
\usepackage{soul}

\title[Thinking Inside the Box]%
      {Thinking Inside the Box: Considerations for Putting Data Physicalization Workshops in a Box}

\author[D. Akbaba, C. Svensson, C. Torelli, M. Callmeryd, \& M. Meyer]
{\parbox{\textwidth}{\centering 
D. Akbaba$^{1}$\orcid{0000-0001-9419-3402}
C. Svensson$^{2}$
C. Torelli$^{2}$
M. Callmeryd$^{3}$
and M. Meyer$^{2}$\orcid{0000-0002-8971-6245}}\\
{\parbox{\textwidth}{\centering 
$^1$KTH Royal Institute of Technology, Sweden\\
$^2$Linköping University, Sweden\\
$^3$Östergötlands Stadsmission, Sweden}}
}

\begin{document}


\maketitle
\begin{abstract}
Visualization researchers utilize workshops both for applied research and to engage different populations with visualization-based activities. While there are many benefits to running visualization workshops, their utility and impact rely on the presence of a researcher who has deep knowledge about visualization theory and practice. In this work, we introduce workshop-in-a-box as a design concept intended to challenge the researcher-centric approach to data physicalization workshops. Through a design study with a socially innovative organization, we deployed several data physicalization workshops that our collaborator ran instead of us. Based on this experience, along with two accompanying case studies that validate the concept, we present material and procedural considerations for how to put data physicalization workshops into a box and the implications it has for extending visualization research outside the bounds of academia.


\end{abstract}  
\newcommand{\artifact}{\raisebox{-.2ex}{\includegraphics[height=1.3\fontcharht\font`\B]{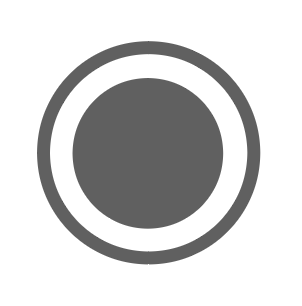}}}

\newcommand{\datacards}{\href{osf.io}{\nearrow data cards}}

\section{Introduction}
Workshops are a common method in visualization research to co-define research goals~\cite{kerzner2019framework, knoll_extending_2020, cay_colvis_2020}, to teach specific topics~\cite{bae2023cultivating, verhaert2021datablokken, kejstova2023construct}, and to generally engage participants in creative processes~\cite{cay2024cobadge,bhargava2017data}. While workshops remain a common approach to reach formative and evaluative research ends, a recent spate of work focuses on workshop formats to facilitate creative interactions between participants and data. These workshops range in topics from data privacy~\cite{verhaert2021datablokken} to critical data literacy~\cite{bae2023cultivating}. A common theme across these workshops is that they employ unplugged and analog activities to lower the barrier to entry. Researchers use these \textit{data physicalization} workshops to engage participants from children~\cite{kejstova2023construct, bae2023cultivating} to adults~\cite{cay2024cobadge, huron2017physical}.

We conducted a design study where we co-designed data physicalization workshops with a socially-driven organization in \sweden{}. Our primary collaborator, a coauthor on this paper, works at the local branch of a national organization called \organization{} --- an organization that runs a variety of programs to support marginalized communities, including vulnerable young adults. As a socially innovative organization, our collaborator wanted to develop new tools that he and his team could use to facilitate meaningful dialogue, self-reflection, and self-expression for their target groups. The iterative co-design process was shaped by pragmatic constraints: a shared goal to use data visualization for social innovation and our collaborator's vision to develop reusable workshop material that anyone across his organization could use. Driven by the twin goals of deploying a tool \textit{and} social innovation, we developed a series of data physicalization workshops piloted with one of \organization{}'s youth leadership programs.

While we found abundant examples of how to create engaging data physicalization workshops, we stumbled on how to make the workshops reusable and easy to facilitate. Our collaborator required workshop material that he and others at \organization{} could run \textit{without} a visualization researcher or deep visualization knowledge. 
Although data physicalization workshops have been successful for a variety of public engagement activities, their documented impact is limited to situations where a visualization researcher is present. 
We began to wonder, what would a deployed version of a workshop look like? And how could others, with limited knowledge of data visualization, run a visualization workshop? 

As we worked iteratively with our collaborator to develop the workshops, our design decisions were shaped by recent literature critiquing the end of design studies. This body of work~\cite{meyer2020criteria, akbaba2023troubling} drew our attention to limitations and unintentional harms caused by researchers suddenly leaving collaborators with half-functioning tools or insufficient support to continue the project independently. As a corrective, we turned our attention to creating sustainable and persistent workshop material. Each iteration of the pilot workshops focused on developing materials and scaffolding that could enable our collaborator and his colleagues to run the data physicalization workshops without us. These design considerations required us to reflectively identify and articulate our visualization and workshop facilitation knowledge. We aimed to put all we knew in a figurative box, hence \textit{\wib{w}}.

In this paper, we introduce \wib{w} as 
a provocative design concept intended to challenge and encourage researchers to reflect on the knowledge they unintentionally exclude from workshop instructions and materials. 
We argue that putting workshops in boxes is important for creating visualization material that are sustainable and continue to support collaborators after the research collaboration ends.
\rev{While we position our work as specific to data physicalization workshops, we speculate that many of the takeaways could apply more broadly to community-oriented work.}
To support scrutiny of our research process~\cite{meyer2020criteria, rogers2021insights}, we present abundant and thick descriptions of the initial pilot material from three workshops that were deployed between January--May 2024 as deep links throughout the paper. Additionally, we validate our method with two further workshop examples that ran independently of the workshop team, and provide suggestions for future visualization researchers to consider when putting their own workshops in a box.
Finally, we reflect on what did not make it into the box, identifying future areas of work essential to push visualization research toward sustainability and accessibility, bridging the gap between academia and communities.
\vspace{-1em}
\section{Related Work}
\label{sect:related-work}
A workshop is a relatively short gathering of people that relies on specific and structured methods to focus on a certain topic or to achieve a particular goal within a shared domain~\cite{kerzner2019thesis, orngreen2017workshops}. As a research method, workshops are employed either in the service of the researcher or in the service of the participants. More specifically, when workshops are employed to assist researchers, they are either formative --- providing new insight into the research problem --- or evaluative --- evaluating a proposed solution. In contrast, workshops in the service of participants are designed research outputs, intended to support a specific outcome, like teaching a topic or spurring creativity. 
In this section, we review workshops used and designed by visualization researchers, and the range of skills required to run them. We also draw attention to the ways that existing work has focused solely on visualization researchers as facilitators, limiting the impact and sustainability of workshops --- a core focus of our \wib{w} provocation.
\vspace{-1em}

\subsection{Workshops for Research} 
Formative workshops are used as a method to gain a deeper understanding of a specific research problem or goal. One example of this is the creative visualization-opportunity (CVO) workshop~\cite{kerzner2019framework}, which intentionally employs the workshop format to reduce the time and resources necessary for visualization researchers to characterize domain problems, analyze needs, and scope visualization opportunities. CVO builds off prior work that considered the need to educate stakeholders about the variety and potential of visualizations~\cite{koh2011developing,goodwin2013creative}. 
The CVO workshop framework contributes guidance for visualization researchers to design, execute, and analyze workshops in collaborative and applied work. 
Continued work by Knoll et al.~\cite{knoll_extending_2020} extends the framework with additional guidelines focusing on supporting interpersonal dynamics and methods for analyzing workshop outputs.
Relatedly, Çay et al.~\cite{cay_colvis_2020} propose a workshop framework that applies user-centered design principles to data visualization workshops, demonstrating how the framework promotes design thinking and collaborative decision-making.

There are several examples of how workshops support formative visualization research efforts.  Such examples include co-creation with social scientists~\cite{leon2020cocreating}, energy supplier analysts~\cite{goodwin2013creative}, and clinicians~\cite{lin2021sanguine}. Seen from these examples and others~\cite{schottler2025practices}, while workshops are effectively used with practitioners, they are not feasible for all populations~\cite{knoll2025gulf}.

Visualization researchers also use workshops to evaluate visualization designs. One example is the workshop run by Bach and colleagues~\cite{bach2022dashboarda} to assess design patterns for dashboards. 
This workshop created opportunities for the researchers to receive feedback from workshop participants about their design pattern collection in an interactive and iterative format. 
In another evaluative workshop study, Knoll et al.~\cite{knoll2025gulf} employ the workshop format in lieu of interviews because the output of collaborative participation is often more productive and varied than individual sensemaking. In their study, the authors utilized workshops to investigate the alignment between intended chart messages and those understood by diverse audiences. 
As a final example, researchers employed workshops to co-design a microgrid to both elicit the design problem \textit{and} to gather feedback from key stakeholders~\cite{ens2021uplift}.

\vspace{-1em}
\subsection{Workshops for Engagement}
While formative and evaluative workshops are important tools for visualization researchers, they are also used for participant engagement in a range of topics.
Numerous examples come from instances where researchers seek to create new methods for education. As an example, Bae et al.~\cite{bae2023cultivating} conducted workshops with children and teachers using their design probe to understand how unplugged methods could support data visualization literacy. 
Other researchers have also used workshops and physicalization to educate children about data~\cite{kejstova2023construct, verhaert2021datablokken}. 
Relatedly, He \& Adar~\cite{he2017vizitcards} designed cards along with a workshop as a structured approach for design activities within classrooms. The card-based workshop introduces game mechanics that bridge visualization theory with practice for university students. In a format that blends evaluation with education, Bach et al. used workshops to test design patterns they developed for teaching data comics~\cite{bach2018design}. In this instance, the workshop provides ecologically valid feedback about the usability in educational environments. 

Many researchers also turn to data physicalization workshops as a method to support community engagement. Bhargava \& D'Ignazio~\cite{bhargava2017data} begin workshops with data physicalization activities to break the ice and help people feel comfortable around data. In a longer-form engagement, Perovich et al.~\cite{perovich2021chemicals} use workshops early in their community-oriented project to engage citizens with open environmental data. Not only did the workshops contribute to trust-building and project ideation with community members, but also led to additional funding to deploy one of the workshop ideas. Additional examples also focus on scaffolding creativity at the start of longer community activities~\cite{cay2024cobadge, dove2014using}. 
In our work, we contribute to these examples of how data physicalization workshops can be used for engagement purposes.

\vspace{-1em}
\subsection{Complexity of Running Workshops}
Although workshops are a widely used method within visualization research, they are not simple to run. Kerzner et al.~\cite{kerzner2019framework} note that effective workshops are \textit{``design problems''} and, in this vein, they provide many pragmatic suggestions for facilitators such as piloting materials and methods, scheduling breaks, and taking time afterwards to reflect and review workshop artifacts. And yet, there is much to learn as a facilitator that goes beyond domain knowledge and workshop preparation. In a study of novice design facilitators, Gu et al.~\cite{gu2023experiences} note that the students recruited as facilitators had trouble with soft skills such as staying comfortable with silence, staying open to unexpected directions and discussion, and cold-calling on participants to participate. This study found, though, that with ample support before and during the workshops --- including an introduction to tools, the researchers' workshop philosophies, and workshop material --- the novice facilitators overcame their discomfort and ran successful remote workshops. Others also emphasize the importance of interpersonal skills when running workshops, like speaking the language of participants~\cite{knoll_extending_2020}, actively supporting participants~\cite{he2017vizitcards}, and keeping the discussion on track~\cite{cay_colvis_2020}. And several note the importance of location in workshop design: both the relevance to the topic and as a space that feels inviting and fosters feelings of equality~\cite{bressa2019sketching, cay_colvis_2020,knoll_extending_2020}.


Acknowledging the difficulty of facilitation, several researchers offer specific advice and guidelines for running workshops. Roberts et al.~\cite{roberts2022reflections} collectively reflect on educational activities for visualization education. They provide guidelines for designing and running such activities, highlighting important considerations of learning goals, workshop design, and facilitation. In contrast, Huron et al.~\cite{huron2016using} published a concise manual for using a data physicalization toolkit. In this instruction guide, the authors provide useful context and other procedural knowledge to enable other researchers to facilitate workshops using the toolkit. They continue in later work~\cite{huron2017physical} with more elaborate facilitator techniques that account for common pitfalls participants face when working with data, ideation, and data physicalization construction. Implicit, however, is the assumption that workshops will be run by researchers. \rev{While there are many known good practices on how the run workshops \textit{as} the facilitator, there is little guidance on how to design for others.} 
\rev{When we sought guidance on this, we found little content aside from material designed for educators~\cite{scottishdataeducation}.}

\vspace{-1em}
\section{Background}
In this section, we describe the domain and research context for our design study with \organization{}, which began in November 2023 and is ongoing. We describe the \textit{specificities} of the research context, intentionally drawing attention to the role of situatatedness and relationality in place of the standard task abstraction as a response to recent work that calls forth the entangled and interpretative nature of visualization design study~\cite{akbaba2024entanglements, meyer2020criteria, tran2024situating}. 

\vspace{-1em}
\subsection{Domain Context}
\organization{} is a nonprofit public benefit organization that works directly in communities across \sweden{} to support marginalized individuals through a variety of programs and initiatives. Their mission is to reduce suffering by working with vulnerable youth, women, immigrant groups, the elderly, and other individuals who are facing economic vulnerability, homelessness, addiction, or poverty. 
\organization{} aims to improve individuals' quality of life by supporting their independence and resilience. 

Broadly, this type of work falls under \textit{social innovation}, which describes organizations that work toward improved social outcomes~\cite{phillips2015social}. In Europe, social innovation has increasingly garnered attention as a shared goal among organizations, governmental agencies, academia, civil society, and the private sector to work toward new and creative ideas that address societal challenges~\cite{eusocialinnovation}. Social innovation initiatives typically target employment, education, social inclusion, and environmental sustainability. Transnational social innovation associations support local organizations, like \organization{}, by providing funding, promoting knowledge sharing, and scaling pilot projects toward the larger aims of improving societal outcomes.

When we began to work with \callmeryd{}, the local \organization{} had just received funding for building up a youth leadership program for teenagers and young adults aged 16 to 25 in two of the city's neighborhoods. 
\callmeryd{} led the program team as they set out to create a series of empowering activities that would strike a balance between being both fun and serious in order to spur discussion and reflection. The program team at \organization{} also focused on connecting the young leaders with different people across the city to expand their networks, build connections, and learn from others, including politicians, business owners, and researchers like ourselves. 
Our collaboration with \organization{} began because of \callmeryd{}'s desire to both socially innovate and to connect the teens with the local university. This resulted in a series of workshops that we co-designed with \callmeryd{} for the leadership program, and that form the basis for our research discussed in this paper.

We applied for approval from the \sweden{} Ethical Review Authority and received an exemption decision. The workshop participants were 15--20 young women aged between 18--21, who were active participants in \organization{}'s leadership program. These young women had created their own leadership group within the program, which they called \pidh{}. We knew that the women were transnational youth with parents and families who came from other countries to \sweden{}. The leadership program supported these young women in finding novel ways to express themselves and in exploring opportunities for further personal development, whether through formal education or entrepreneurship. Many of the young women were highly motivated to participate in the program, but like teenagers anywhere, there were also some with less motivation. Although we observed several different friend cliques, the general atmosphere was congenial and supportive. 
Aside from the background on where their families came from and their current neighborhoods, we knew very little about their lives, educational background, and personal goals. This led us to rely heavily on \callmeryd{} for feedback on the appropriateness of the activities we were co-designing together.


\vspace{-1em}
\subsection{Research Context}
The research team comprised two visualization researchers --- \akbaba{}, a PhD student at the time, and \meyer{}, her supervisor --- and two Masters of Design students, \torelli{} and \svensson{}. 
The majority of the pilot workshops were designed, iterated, and evaluated by \torelli{} and \svensson{} for their joint Master's thesis, with \akbaba{} and \meyer{} providing feedback and guidance along the way.

As a research team, we held a specific epistemic stance that shaped how we approached the research. We share this stance here due to recent calls across visualization~\cite{akbaba2024entanglements} and other scientific domains~\cite{jamieson2023reflexivity} to increase transparency in research practices by providing important context for how researchers design, conduct, and make sense of their research. Broadly, we consider alternative epistemic perspectives in visualization research productive and insightful. As a result, many of our decisions were guided by principles found across data feminism~\cite{dignazio2023datafeminism}, entanglements~\cite{akbaba2024entanglements}, data humanism~\cite{lupi2017data}, feminist care ethics~\cite{puig2011matters}, and critical data studies~\cite{gitelman2013raw, loukissas2019all}. Taken together, these principles shaped a set of values that drew our attention to certain aspects of the collaboration and design process over others. Explicitly, this meant that we aimed to create workshops that demonstrated how data can be personal and meaningful, even when small, subjective, and open to interpretation. This stance views data as produced rather than as objects found in the world.
\vspace{-0.5em}

Additionally, to align our design study with our collaborator's twin goals of empowering people and leaving a lasting impact, we actively engaged in an alternative epistemic approach to design study. As others have noted~\cite{meyer2020criteria, akbaba2023troubling}, the original design study methodology~\cite{sedlmair2012design} does not offer guidance on how to ethically exit and account for the lasting impacts of working together with others. For this reason, we conducted our collaboration as a \textit{care-ful design study}~\cite{akbaba2023troubling}. This means that throughout the process, our design decisions were not only informed by the tasks at hand, but also with a consideration for what could happen after we inevitably \textit{exited} the collaboration. Furthermore, our collaborator's focus on empowerment --- as opposed to the typical goal of analysis and insight --- challenged our ideas of visualization and data within a design study context. 
\vspace{-0.5em}

Given the age of the participants and the general goals of the leadership program, we decided to create a series of data physicalization workshops due to their documented successes in engaging a wide range of audiences. \rev{Prior work demonstrates that data physicalizations have a lower barrier of entry in terms of visualization literacy, support active engagement, and spur self-reflection~\cite{huron2014constructing, bae2022making, thudt2018self}}. The physicalization activities were drawn from recent academic publications that highlight the advantages of using unplugged, or non-digital, methods to teach concepts important to data visualization~\cite{bae2023cultivating, bhargava2017data, koningsbruggen2022what, hogan2017pedagogy, augcca2023data} as well as more creative endeavors~\cite{dragicevic2020data, warnes2018}.
Specifically, we developed crafting exercises that would be fun and engaging for the young women by focusing on creative activities that can support self-expression. We used material that was easily found in local shops to emphasize that these activities could also be done at home if the women wanted.

\vspace{-0.5em}
To support reflection and discussion, we turned to the \textit{Dear Data}~\cite{lupi2016dear} project. Dear Data was a year-long collaboration-turned-book of two designers, Lupi and Posavec, who mailed each other hand-drawn postcards visualizing a week of collected data on some topic, highlighting the ways that small, mundane datasets can tell nuanced and insightful stories. We found the data stories shared between the authors inspiring, beautiful, and incredibly rich, conveying the textures of their daily lives. Their correspondence demonstrated that mundane data are powerful drivers for storytelling and that visual methods can break standard minimalist conventions and \textit{still} maintain legibility to others.  We suggested a series of workshops that would guide the participants through a similar process of collecting personal, mundane data and then turning it into a non-digital data visualization.

And finally, we decided \textit{not} to facilitate the workshops for several reasons. We encountered the pragmatic constraints of a language barrier between the research team and the young women, and a need for long-term trust necessary to have reflective discussions. For these reasons, we decided \callmeryd{} would be the best facilitator among us. \callmeryd{} also shared a strong desire to develop methods that anyone across \organization{} could use, which echoed our considerations regarding the sustainability and ethical exit of our collaboration. These constraints led us to consider how to put our workshops --- with all the associated presentations, instructions, visualization examples, and activities ---  in a metaphorical box so that others could facilitate similar data physicalization workshops with minimal training and no access to a visualization researcher.

\newcommand{\inlinebox}[3]{
    \fcolorbox{#1}{#2}{#3}
}

\vspace{-1em}
\section{Pilot Workshops}
We developed a series of three data physicalization workshops, each three hours long. The workshops were held in the science center connected to the university (twice) and the headquarters for the youth leadership program (once). Each workshop followed a similar format: there was a central question that would be posed to the group; they would create personal datasets about the topic, followed by a data physicalization activity; and the workshop concluded with a discussion and group reflection about the topic using the data and physicalization artifact as a discussion point. 

The workshop material was iteratively developed by the research team working with \callmeryd{}. \torelli{} and \svensson{} worked closely with \akbaba{} and \meyer{} to develop initial activity ideas. Before each workshop, we held iterative meetings with \callmeryd{}, and occasionally some of his colleagues, to reach a point of \textit{task clarity}~\cite{sedlmair2012design} about the upcoming workshop's format and focus. These meetings also helped \callmeryd{} to familiarize with the material and become comfortable with the data physicalization activities.
During the workshop, as \callmeryd{} facilitated, part or all of the research team would attend to take field notes and offer assistance if needed.
Afterward, we shared our notes from the workshop and collected informal feedback from \callmeryd{} through semi-structured interviews. The feedback from interviews and our group discussions shaped our iterative design of subsequent workshops. 

In the following sections below, we provide descriptions of the workshop aims and activities, with further details presented in \torelli{}'s and \svensson{}'s joint thesis~\cite{svensson2024visualizing}. 
Additionally, we provide deep links to the workshop material for supporting traceability~\cite{rogers2021insights}, what we included in the figurative boxes --- the  \inlinebox{boxbordercolor}{boxcolor}{materials} and the \inlinebox{boxbordercolor}{boxpcolor}{procedures} for transferring knowledge  ---
and reflections on whether the box was enough for \callmeryd{} to run a data physicalization workshop on his own.

\vspace{-1em}
\subsection{Pilot 1. Data Visualization Warm-Up with \textit{Data Selfies}}

For the first workshop, we were excited and eager to test out the ways that visualization can be used as a method of self-expression. We wanted to create an empowering workshop that was fun and engaging for the young women and offered a different medium of self-expression à la \textit{Dear Data}. We felt that it was particularly important to offer many opportunities for creativity so that the young women could personalize their visualization because of the stereotyping they receive in society. However, given the range in educational backgrounds and the age of the participants, we decided that first we would need to introduce the concepts of data and the principles of visualization in a quick and easy way. This light introduction also supported subsequent workshops as foundational knowledge.

To accomplish this goal, we adopted the Dear Data lesson plan developed by Data Education in Schools~\cite{scottishdataeducation} to quickly introduce the concepts of data and visualizations. While there was a lot of educational material that we would have liked to include in the workshop, our choices of what to include were bound by the reality that we only had three hours with the young women, who would also be tired when attending the workshop after school. 

\begin{figure}
    \centering
    \includegraphics[width=1\linewidth]{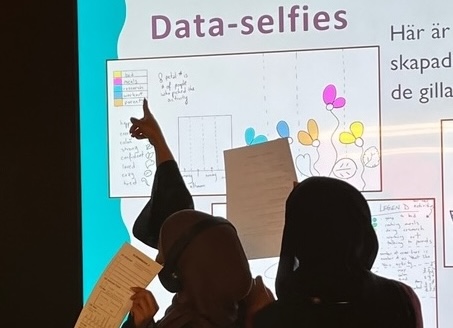}
    \caption{Workshop participants from the first pilot, \textit{Data Selfies}, using a presented example to make sense of the activity.}
    \label{fig:pilot-pointing}
    \vspace{-10pt}
\end{figure}

The first workshop, titled \textit{Data Selfies} and shown in Fig. \ref{fig:pilot-pointing}, balanced these constraints with our goal to introduce data visualization as a method for self-expression. The first half of the workshop introduced the concept of data and datasets. As a group, the young women collectively created a group dataset about the activities they did in their free time. We focused on activities because we wanted a dataset that was easily tokenized, mundane, yet also personal enough to reflect the unique interests of the participants. The second half of the workshop presented visualizations as a tool for self-expression. To support creativity, we provided blank legends, instructions, and craft material that would translate their data into thematic visualizations --- like \href{https://osf.io/sxv5w/files/zb2f3}{$\nearrow$ flowers} or \href{https://osf.io/sxv5w/files/eshfa}{$\nearrow$ houses}. The workshop box consisted of the following materials:

\wibbox{
\textbf{The Data Selfie Materials Box}
\begin{itemize}
    \item \href{https://osf.io/sxv5w/files/y28qp}{$\nearrow$ presentation} introducing data and visualization 
    \item \href{https://osf.io/sxv5w/files/24fyv}{$\nearrow$ data collection cards}
    \item giant printed \href{https://osf.io/sxv5w/files/aus3r}{$\nearrow$ table}
    \item crafting materials (e.g., markers, origami paper, stickers, scissors, glue)
    \item \href{https://osf.io/sxv5w/files/j7cpz}{$\nearrow$ facilitator guide}
    \item blank legends and instructions for \href{https://osf.io/sxv5w/files/zb2f3}{$\nearrow$ flower}, \href{https://osf.io/sxv5w/files/eshfa}{$\nearrow$ house}, \href{https://osf.io/sxv5w/files/ywqrz}{$\nearrow$ hot air balloon},
    \href{https://osf.io/sxv5w/files/ak478}{$\nearrow$ dresses},
    or \href{https://osf.io/sxv5w/files/9muyv}{$\nearrow$ abstract art} data visualizations
\end{itemize}
}

Along with the materials, we had several processes to support \callmeryd{} in his role as facilitator. We spent a considerable amount of time walking through the material prior to the workshop so that he felt comfortable talking about data visualization concepts. These meetings not only covered the \href{https://osf.io/sxv5w/files/y28qp}{$\nearrow$ presentation} material, but also a step-by-step \href{https://osf.io/sxv5w/files/j7cpz}{$\nearrow$ facilitator guide} that outlined what he and his co-facilitator would do during each activity. This allowed \callmeryd{} to understand the workshop mechanics without having run it before. 

\proceduresbox{
\textbf{The Data Selfie Procedures Box}
\begin{itemize}
    \item pre-workshop meetings to review visualization concepts
    \item email exchanges for refining workshop materials
    \item designing a \href{https://osf.io/sxv5w/files/j7cpz}{$\nearrow$ facilitator guide}
\end{itemize}
}

Unfortunately, the contents of the box were not enough. The second half of the workshop, with the data visualization activity, resulted in more frustration than visualizations. Out of three groups, only one was able to create a partial data visualization, while the other two were left upset and confused. 
In our attempt to optimize self-expression, we underestimated the difficulty of going from a dataset to a designed legend. We had not prepared \callmeryd{} on how to provide adequate support during the activity, and the instructions --- which included the \swedish{} terms about data, legends, and visualization --- were unfamiliar for some of the participants. We abandoned the activity and had a debrief before the workshop ended to reassure the participants that the lack of success was our fault and not theirs.
We used these setbacks as discussion points in a productive post-workshop discussion with \callmeryd{} regarding what to change for the next workshop.

\vspace{-1em}
\subsection{Pilot 2. Simple Data Collection with \textit{Data Jewelry}}
\begin{figure}
    \centering
    \includegraphics[width=1\linewidth]{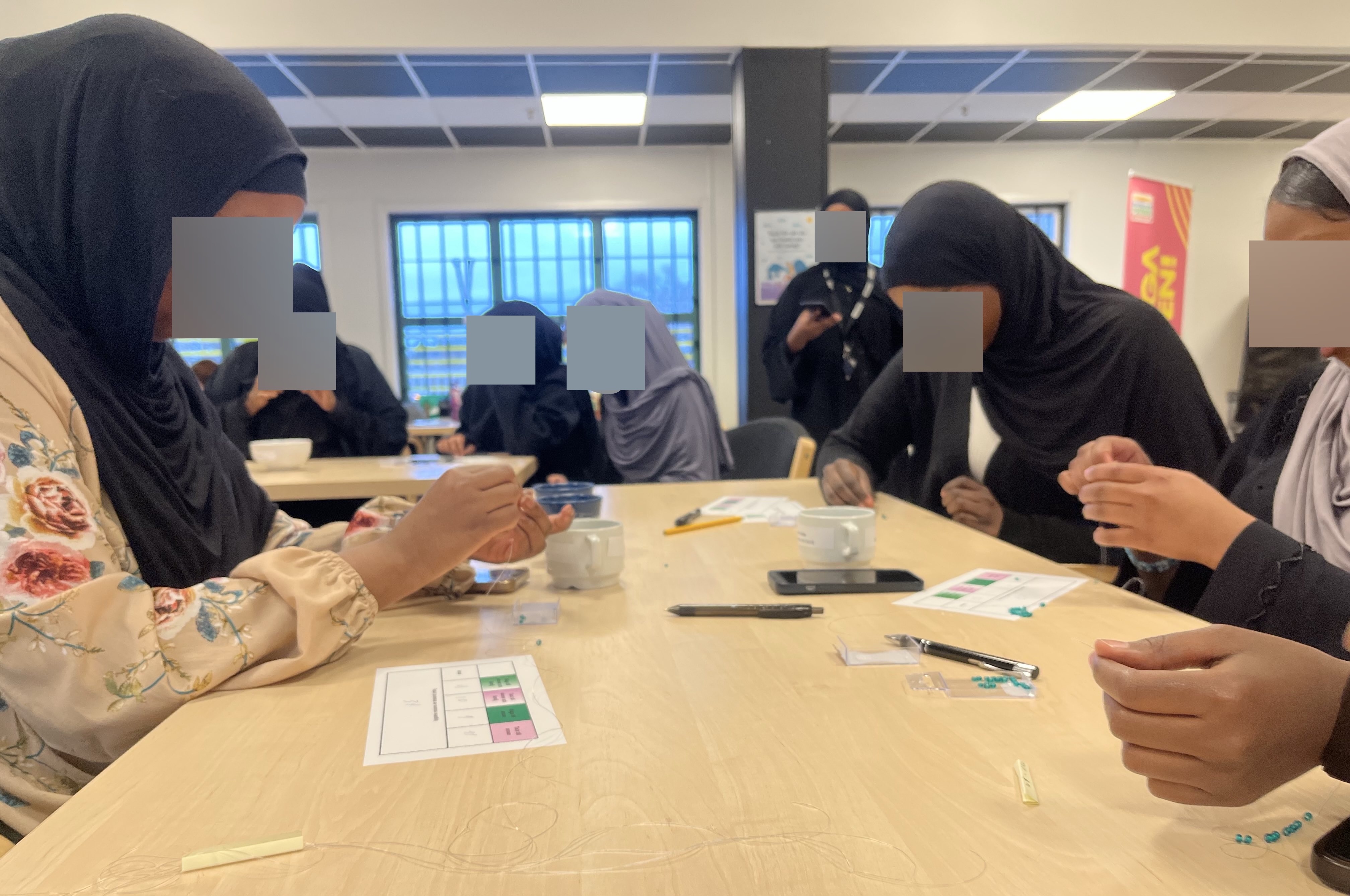}
    \caption{Participants from the second pilot, \textit{Data Jewelry}, using different colored beads and string to represent data they collected over four days, where they tracked the frequency of complaints and utterances of gratitude.}
    \label{fig:data-jewelry}
    \vspace{-15pt}
\end{figure}

Despite the challenges with visualizing the data, the first workshop demonstrated how the process of co-creating datasets could be accessible and lead to reflective and insightful discussions for \callmeryd{}. We maintained the design goals of supporting creativity and self-expression, but we shifted our attention to the workshop as a tool to spark meaningful conversations between \callmeryd{} and the young women. When we began to brainstorm topics for the workshop, however, we had a very difficult time deciding on what data the young women should and could collect. There was a clear gap in expertise, experience, and knowledge that made it difficult to come up with a solution. On the research team, we knew too little about the young women and \organization{}'s goals to offer suggestions. On the other side, \callmeryd{} suggested broad topics of interest, like how the young women can feel empowered across multiple scales of injustice, that were important to \organization{} but incredibly difficult to tokenize given the time constraints of the workshops. Our answer came from flipping through \textit{Dear Data} and finding inspiration in their week three postcard, where they visualized all the times they said thank you. We decided that the women would collect data about the number of times they expressed gratitude or complained.

For the purpose of supporting dialogue, we posited that richer data, which came from \textit{in situ} recording rather than \textit{post hoc} generation, would spark a more interesting conversation because the data would better reflect the lives of the young women. 
For four days, the young women recorded when they expressed gratitude or complained to either a friend or family member and similarly to a stranger --- creating four data items to keep track of --- \{1: Gratitude--Friends/Family, 2: Complaint--Friends/Family, 3: Gratitude--Stranger, 4: Complaint--Stranger\}. We chose a simple phone app that was a \href{https://osf.io/sxv5w/files/bd3t9}{$\nearrow$ tally system} with a nice UI. We wanted the app to be straightforward to use and accessible on all operating systems. We instructed the group to set a timer for every two hours, in case they forgot to record in situ.

For this workshop, we also decided to change the visualization activity so that it was easier to map the data to the visual abstraction. Each collected data item was tokenized as a unique physical representation in the form of a bead. During the workshop,  \callmeryd{} guided the group through the process of creating data jewelry from the data they had collected over four days, shown in Fig. \ref{fig:data-jewelry}. The \href{https://osf.io/sxv5w/files/8dswq}{$\nearrow$ four different beads} corresponded to the data types that had been collected. Afterward, \callmeryd{} led a discussion about how the group perceives their lives through the lens of gratitude or complaints.
\vspace{-0.5em}

\wibbox{
\textbf{The Data Jewelry Materials Box}
\begin{itemize}
    \item pre-workshop \href{https://osf.io/sxv5w/files/bd3t9}{$\nearrow$ instructions} for data collection
    \item pre-workshop \href{https://osf.io/sxv5w/files/bd3t9}{$\nearrow$ presentation} for data collection
    \item \href{https://osf.io/sxv5w/files/5duxw}{$\nearrow$ presentation} introducing data physicalization
    \item \href{https://osf.io/sxv5w/files/bjx42}{$\nearrow$ facilitator guide}
    \item \href{https://osf.io/sxv5w/files/d76wa}{$\nearrow$ data collection cards}
    \item jewelry material (beads, string)
\end{itemize}
}

\vspace{-0.5em} 
A crucial step to this activity's success was setting up the data collection app ahead of time. We tested the app beforehand by conducting our own data collection exercise. The exercise demonstrated that we needed to set up a reminder timer because, despite the intention to record data, it was not always easy to remember throughout the day. After piloting the data collection, \torelli{} and \svensson{} made an \href{https://osf.io/sxv5w/files/bd3t9}{$\nearrow$ instruction manual} that \callmeryd{} reviewed with the young women. In this pre-workshop activity, \callmeryd{} helped the young women set up the app and the reminder, and discussed what types of data they could collect over the next four days. 

\proceduresbox{
\textbf{The Data Jewelry Procedures Box}
\begin{itemize}
    \item piloting the data collection activity together before the workshop
    \item pre-workshop meeting to review workshop content
    \item choosing specific materials for the jewelry
\end{itemize}
}

While all of the young women collected data over the four days, the amount of data they had collected greatly varied. This resulted in some data jewelry that barely constituted the circumference of a ring, and others that filled a long necklace. We intentionally did not require a minimum number of data points because we wanted the data collection exercise to be accessible to all the participants. And yet, this posed an interesting challenge for workshop planning in terms of ordering material, timing, and creating a wearable output. 

Furthermore, the activity ended very quickly. Not only was there less data to work with for most of the young women, but the data mapping task was greatly simplified by pairing unique physical tokens with unique data categories. Despite this, \callmeryd{} filled the time with more discussion with the group using the categories of collected data as the starting point for deeper conversations about how family and cultural values influence the ways the women paid attention to how and to whom they expressed gratitude. In our post-workshop discussion with \callmeryd{} he commented that the deeper conversations --- as facilitated by the data physicalization activity --- were a successful outcome for the workshop. 

\vspace{-1em}
\subsection{Pilot 3. Mixed Media Data Collection with \textit{Data Mapping}}
\begin{figure}
    \centering
    \includegraphics[width=1\linewidth]{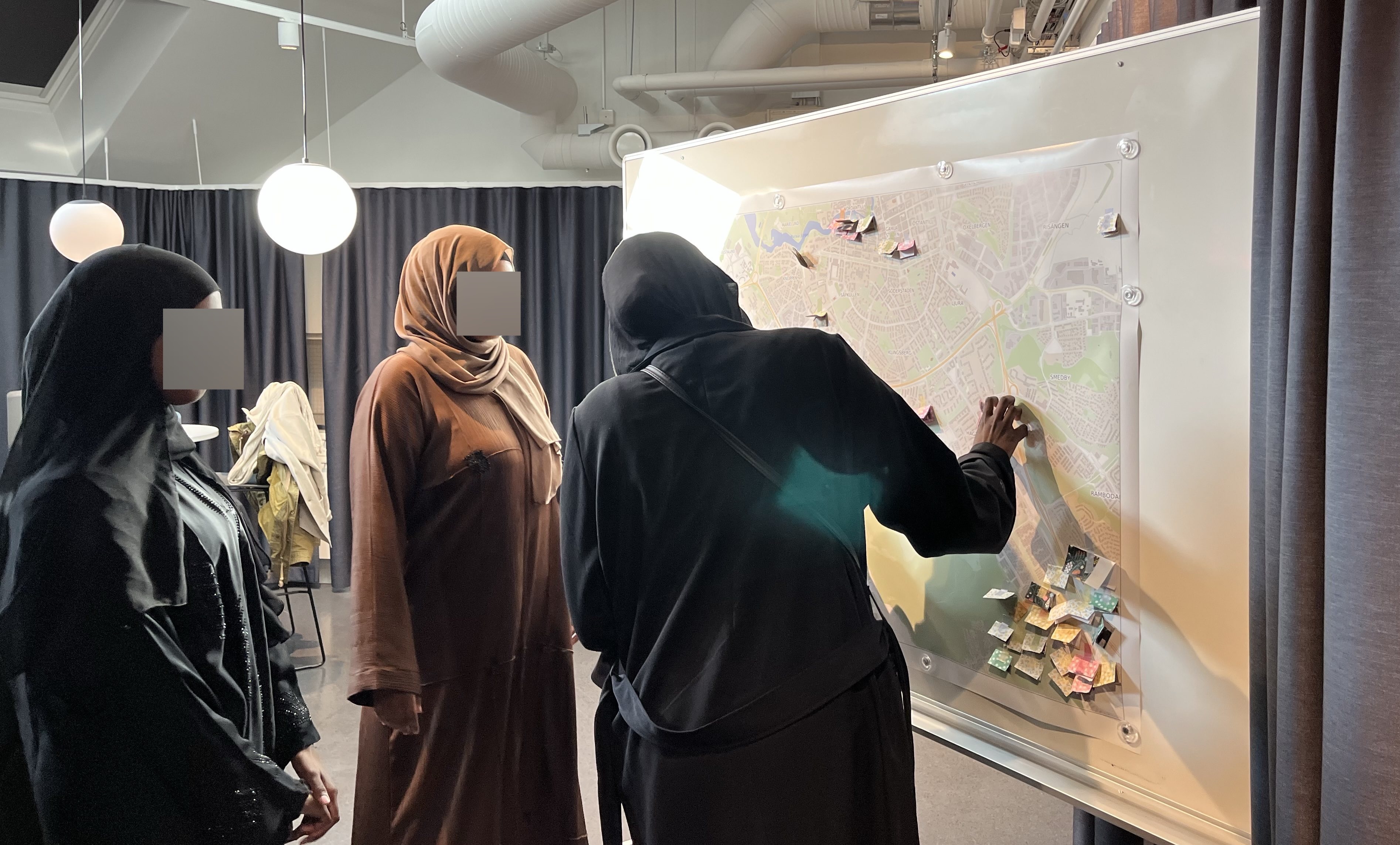}
    \caption{After the data physicalization exercise in the third pilot, \textit{Data Mapping}, workshop participants reflected on the map created during the workshop.}
    \label{fig:data-mapping}
    \vspace{-15pt}
\end{figure}

In the final workshop, we aimed for more consistency in data collection across the workshop participants because we saw the potential of using data physicalization workshops as a tool for \callmeryd{} and \organization{} to spark interesting discussions with target audiences. The second workshop demonstrated that there was a wide range in \textit{how much} data the participants collected, affecting the data physicalization outputs. Thus, for this workshop, we standardized the data collection.

Again, we stumbled on what question to ask of the group. We had not yet found a clear way to operationalize socially-oriented questions into easily tokenized data. We settled on the topic of how the young women felt and moved around the city. 
In later feedback, \callmeryd{} noted that it would have helped to bring a group of staff together from \organization{} in the start of the collaboration for creating
a question bank that operationalized empowerment. 

We also deliberated about whether we should lean on more involved forms of data creation, like a daily diary, or something easier, like a questionnaire, because we wanted to respect the limited time that these young women had in their busy lives. After several design iterations with internal testing, we felt that the best data visualization results would come from an easy collection method with varied data types.  To standardize the data collection, \organization{} created a survey questionnaire --- a format the young women were familiar with --- that they would fill out at four random times. The \href{https://osf.io/sxv5w/files/cej3b}{$\nearrow$ questionnaire} collected a snapshot of the participants' day: what they saw in front of them (photo), who they were with (text), how they felt at the moment (category), and a short description of where they were (text). We planned a backup activity in case participants did not fill out the survey or only submitted partial survey answers. Our goal was to create a map of their experiences in the city during the physicalization workshop, shown in Fig. \ref{fig:data-mapping}.

Due to time constraints within the leadership program schedule, the young women were emailed a \href{https://osf.io/sxv5w/files/2q9u6}{slide deck} with instructions for the data collection. Without a meeting, the young women did not get the opportunity to express their qualms about submitting photos, which we only learned about during the workshop. Additionally, a few had not read the instructions and did not set up reminder alarms.
As a result, the majority of the women submitted partial or no responses to the four data collection questionnaires. 

\wibbox{
\textbf{The Data Mapping Materials Box}
\begin{itemize}
    \item pre-workshop \href{https://osf.io/sxv5w/files/2q9u6}{$\nearrow$ instructions} for data collection
    \item \href{https://osf.io/sxv5w/files/gx8eb}{$\nearrow$ facilitator guide}
    \item \href{https://osf.io/sxv5w/files/4dbnu}{$\nearrow$ presentation} introducing maps as visualizations
    \item giant printed \href{https://osf.io/sxv5w/files/rwmcq}{$\nearrow$ map} of the city 
    \item partially filled in \href{https://osf.io/sxv5w/files/95qmv}{$\nearrow$ legend} 
    \item crafting material (origami paper, washi tape, stickers)
\end{itemize}
} 

During the workshop, for participants who submitted photos, they glued printed versions of their photos inside a small piece of origami paper that encoded their feelings at the moment. This composite paper was folded over and taped onto a 1 x 1.5m \href{https://osf.io/sxv5w/files/rwmcq}{$\nearrow$ map} 
of the city using colored washi tape to encode who they were with when the photo was taken. Finally, using crosses and hearts, the young women marked places on the map that they avoided visiting (cross stickers) and places they felt welcome (heart stickers).
Because the majority of the workshop participants did not have data to work with, we turned to our backup activity to foster group participation in the activity. Instead of using printed photos and data from the questionnaire, they drew photos and answered the questions post-hoc during the workshop. Even though these participants did not have original data to work with, they still enjoyed the process of contributing to the map. The workshop concluded, following the prior workshop formats, with \callmeryd{} leading a discussion about the data and resulting map.

\proceduresbox{
\textbf{The Data Mapping Procedures Box}
\begin{itemize}
    \item pre-workshop walkthrough of the data visualization \href{https://osf.io/}{activity}
    \item pre-workshop meeting to review workshop content
\end{itemize}
}

While the data collection was less than ideal, we were happy to see that the young women demonstrated comfort with visualization components that they initially found challenging in the first workshop. For example, the map had a partially filled-in legend, with some missing components. The young women easily constructed the second half of the \href{https://osf.io/sxv5w/files/95qmv}{$\nearrow$ legend} after initial instructions. We also noticed that the shyest girls were able to carve space for themselves through data collection, speaking through their volume of data collection, when it was difficult to speak otherwise among louder group members. To our surprise and relief, despite a lack of data from the questionnaires, the participants remained engaged and interested in the physicalization process. It was also clear to us that by the end of the third workshop, \callmeryd{} was very comfortable leading data physicalization activities. Over the course of co-designing data physicalization workshops with \callmeryd{}, we had begun to streamline not only the physical materials he needed, but also the procedures to support his facilitation, \textit{i.e.} we were learning how to put our workshops in a box.

\vspace{-0.5em}
\section{Putting Knowledge in a Box}
\label{sect:wib}

\textit{\wib{W}} is a concept and provocation that we put forward to challenge the researcher-centric approach to designing visualization workshops. Our provocation stems from the constraints of designing data physicalization workshops for our collaborators to facilitate and the joint desire to create sustainable and lasting tools. \rev{Additionally, we clarify that \wib{w} is not a toolkit. As defined by Ledo et al.~\cite{ledo2018evaluation}, toolkits make it easier to ``create new artifacts'' implying something that is ready to use and re-use. \wib{W} centers \textit{processes} for trying to toolkit-ify visualization workshops; it is these processes of working with non-visualization experts that we want to draw our reader’s attention to as a critical moment for reflection around workshop design.} In this section, we detail the parts of \wib{w} that were the trickiest to put in a box, the tacit knowledge surrounding visualization, physicalizations, and workshop facilitation. 


As previously discussed in Sect.~\ref{sect:related-work}, designing, running, and analyzing workshops require facilitator preparation and prior knowledge. While prior work has named these challenges, it has also assumed that the facilitator has \textit{some} visualization background, either as a researcher or practitioner. We identify this as \textit{tacit knowledge} in workshop facilitation. Tacit knowledge refers to the unwritten knowledge that is necessary to complete a task to a certain degree. It has been noted as an important component in scientific endeavors~\cite{collins2001tacit} that is not reported either because it is deemed too fine of detail (e.g., the material of an instrument) or a lack of space for reporting.  And so, our procedural recommendations for \wib{w} are oriented around different ways to transfer knowledge between the visualization researcher designing the workshop and the person who will facilitate it.

\textbf{Strategy \#1. Transferring the knowledge}
A rather straightforward way to share knowledge with future facilitators is to discuss content with collaborators through a series of informational and dialogic meetings. With the initial pilot workshops, we met with \callmeryd{} a few times before each new workshop, with the frequency decreasing as he became increasingly comfortable with data visualizations and physicalization activities. In these meetings, we walked through the workshop's structure and material. This type of information included walking through the different types of assets that would be available to him and his team, as well as the content in the slide decks. We used email to do this exchange asynchronously.

We recommend coming to these meetings with the workshop material prepared beforehand. Creating a slide show, for example, not only structures the meeting but also offers clear entry points for future facilitators to reflect upon, engage with, and discuss the new concepts with the visualization researcher. For example, in our meetings with \callmeryd{}, he asked for more context about data visualization and often asked for additional speaker notes or slides that would help trigger his memory when presenting the material during the workshop.
This type of knowledge transfer is particularly useful for explicating the knowledge necessary to introduce visualization concepts and definitions to workshop participants.
We recommend going through the material in fine detail, scheduling plenty of time for the facilitator to ask questions, request modifications, and provide feedback on other information they might want or need.
\vspace{-2pt}

\textbf{Strategy \#2. Building the knowledge}
Another method of transferring knowledge can come directly through the \textit{experience} of building a data physicalization. We employed this strategy in the second and third workshops --- either collecting data (Data Jewelry) or making the physicalization (Data Mapping) --- to give \callmeryd{} a better feel for the activity. Thus, we recommend running the workshop with the future facilitator as a participant. This activity offers many opportunities to share visualization knowledge. Running through the workshops demonstrates \textit{how} a facilitator can introduce physicalization activities and then support participants while they construct the visualization. By building the data physicalization and listening to instructions, it opens up the opportunity for future facilitators to clarify unclear instructions and to get familiar with the physicalization materials. The process also exposes opportunities to better design the activity. For example, in one of the workshops (Data Jewelry), we identified the need for setting reminder alarms for data collection since some of our team members forgot to collect data. This insight came only after walking through the entire activity together.
\vspace{-.5em}

\textbf{Strategy \#3. Embedding the knowledge}
Embracing the idea that some knowledge comes from trial and error, with this strategy we highlight the types of knowledge that are either difficult to reproduce in smaller workshop settings (Strategy \#2) or too tedious to review along with the other amounts of instructional content (Strategy \#1). For example, for those who have run workshops, it becomes tacit knowledge how to set up a room for effective group dynamics, or the minimum number of scissors per person to reduce bottlenecks. While, of course, you could review these details with the future facilitator, it will likely get in the way of more important information that you are trying to communicate. Therefore, we recommend including some of this tacit knowledge in the design of the workshop. We incorporated this knowledge in every activity that we designed --- from the design of legends (all workshops), to the string material (Data Jewelry), to decisions regarding the size of the map and images (Data Mapping). All of these decisions embody our collective knowledge of designing effective data visualizations and physicalizations.
\vspace{-1em}

\section{Validation}
In order to validate our ideas around \wib{w}, we designed and deployed two separate data physicalization workshops after the initial pilots. In this section, we detail how putting workshops in a box supported two additional facilitator groups.


\vspace{-1em}
\subsection{Case Study 1: Youth Leadership Workshops}
\begin{figure}
    \centering
    \includegraphics[width=1\linewidth]{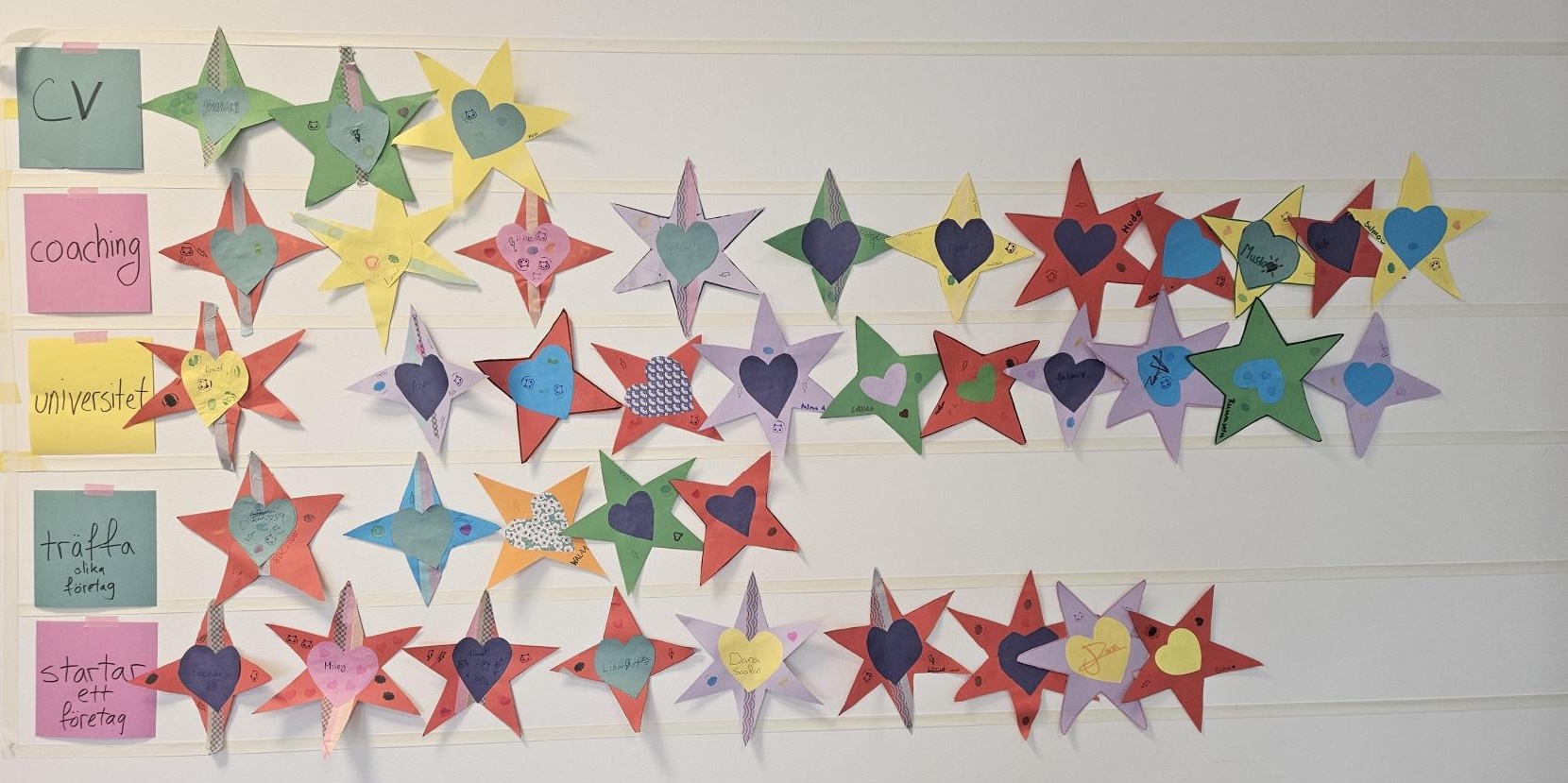}
    \caption{Final group visualization comprised of individual stars created by workshop participants. The stars encode information about the particapants and their dreams for the new year.}
    \label{fig:allstars}
    \vspace{-10pt}
\end{figure}
Following the three pilot workshops, we designed a new workshop with \callmeryd{} to support a discussion about new year's resolutions with another youth leadership program. Titled, \textit{Wish Upon a Star}, see Fig.~\ref{fig:allstars}, this workshop used data physicalization activities to construct a star using personal responses to questions like: \textit{I want support from \organization{} with ...} and encoded responses like: \textit{Getting into university; Starting a business; Meeting business owners}.

The first workshop, run by \callmeryd{} in January 2025, was successful in engaging the youth and providing valuable information to \organization{} about how they should allocate resources for the upcoming months. Given the success of the workshop, another colleague at \organization{} decided that she also wanted to use the workshop with a group of young women, who were members of a separate youth leadership program. In addition to providing the workshop material, \callmeryd{} and \akbaba{} created a five-point overview to provide to the colleague, to summarize the important goals of the workshop. This process of summarizing together offered an opportunity for us, the research team, and our collaborator, to synthesize tacit and practical knowledge together. In other words, \callmeryd{}, who once received a workshop in a box, now contributed to putting the workshop in a box.  We shared this information via email and met once to answer any questions that his colleague might have.

In addition to \href{https://osf.io/sxv5w/files/wa26z}{$\nearrow$ slides}, a \href{https://osf.io/sxv5w/files/pnz2e}{$\nearrow$ legend} for the physicalization, and craft material, our new \wib{w} included high-level facilitator tips that added a \href{https://osf.io/sxv5w/files/gfb7p}{$\nearrow$ meta-overview} on how to prepare as a facilitator. We included pointers like what type of information to share with the workshop participants and how to use the data physicalizations to spark discussion. These meta-points summarized the purpose of the data physicalization workshop as a method for sparking discussion and reflection with the workshop group.

The colleague ran the workshop in February 2025, without \callmeryd{} or anyone from the research team. She reported a successful workshop both in terms of ease in facilitating as well as outcomes from the workshop. The outcomes of the project signaled two things to us. First, that after some time, \callmeryd{} was able to translate his own facilitator's knowledge as additional material for the \wib{w}. This occurred as a result of multiple evaluative discussions after the pilot workshops, where we discussed as a team whether the workshops were truly in a box or not. Second, without the help of \callmeryd{} or the research team, another colleague at \organization{} was able to facilitate the same workshop without having participated in the initial workshop. These stars have been used as conversation starters with public figures, politicians, and others curious about the program and \organization{}'s impact.

\wibBOX{
\textbf{Wish Upon a Star Box}
\begin{itemize}
    \item \href{https://osf.io/sxv5w/files/wa26z}{$\nearrow$ presentation} introducing the activity
    \item material list
    \item \href{https://osf.io/sxv5w/files/pnz2e}{$\nearrow$ legend} for the data physicalization
    \item \href{https://osf.io/sxv5w/files/gfb7p}{$\nearrow$ meta-overview} of activity
    \item examples of pre-made stars
\end{itemize}}{\begin{itemize}
    \item pre-workshop meetings with \callmeryd{} and new facilitator
\end{itemize}}


\vspace{-1em}
\subsection{Case Study 2: Science Center Workshops}

\begin{figure}
    \centering
    \includegraphics[width=1\linewidth]{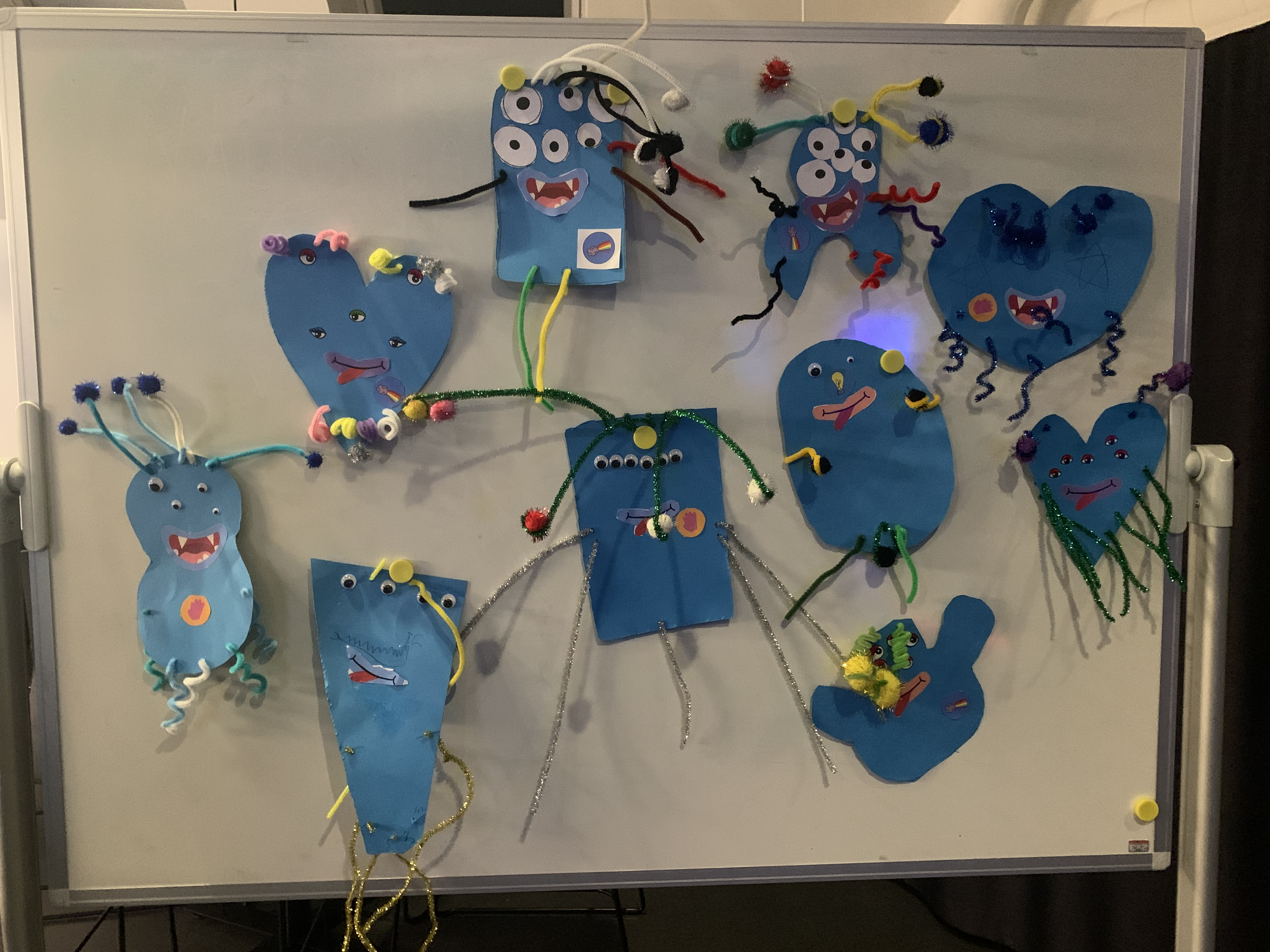}
    \caption{Data aliens that were crafted by school groups. The different body parts and badges represent data collected by the kids as they explored the science center on a guided scavenger hunt.}
    \label{fig:data-aliens}
    \vspace{-17pt}
\end{figure}

After the pilot workshops, we pitched the idea of data physicalization workshops to our local science center,the \visc{}. 
The \visc{} works closely with researchers within our department to bring visualization technology to the public. To accomplish this, it employs engineers, designers, and educators who work with researchers to develop engaging content for children and adults. 
We worked with the educators at the \visc{} to adopt the data physicalization workshops for school groups during a holiday break. These school groups include children ranging from the ages of 7 to 10 years old. 

The workshop design went through a similar process as the pilot workshops, using \wib{w} considerations to develop appropriate scaffolds and instructions for the future workshop facilitators. We developed a data physicalization activity that would utilize answers from a \href{https://osf.io/sxv5w/files/fzj3p}{$\nearrow$ scavenger hunt} throughout exhibits in the \visc{} to build up a data alien, Fig.~\ref{fig:data-aliens}. The workshops ran for several days over the course of a week in February 2025 during a school break. At the end of the week, we met with the educators to gather their feedback.

\wibBOX{
\textbf{Data Aliens Box}
\begin{itemize}
    \item \href{https://osf.io/sxv5w/files/fzj3p}{$\nearrow$ instructions} for data collection and physicalization
    \item \href{https://osf.io/sxv5w/files/h7nu5}{$\nearrow$ printed alien body parts} and badges
    \item \href{https://osf.io/sxv5w/files/g26mh}{$\nearrow$ printable aliens} to mark scavenger hunt clues
\end{itemize}}{\begin{itemize}
    \item pre-workshop alien building activity
    \item pre-workshop meetings to review instructional material
\end{itemize}}

After this workshop, the research team got busy and began to work on other projects. Unbeknownst to us, the educators continued to use the workshop template to create data physicalization workshops for school groups. To our surprise and delight, we learned that they had created a \href{https://osf.io/sxv5w/files/9zspw}{$\nearrow$ data jewelry workshop} for another school holiday in April 2025. Using the data physicalization idea from the second pilot workshop and the questions about activities from the first pilot workshop, the educators modified the workshop to be suitable for young kids by adapting the materials to ensure that the bracelets would look nice and be easy for younger kids to use. Most recently, the educators also adopted the data aliens for a spooky holiday theme, using the idea of the \href{https://osf.io/sxv5w/files/asdne}{$\nearrow$ scavenger hunt} to build up a dataset for ghosts. 
\vspace{-1em}

\section{Discussion: The Things that Didn't Make it Into the Box}
~\label{sect:reflections}
We began our design study with \organization{} with the intention to develop data physicalization workshops for a social innovation setting. The workshops were co-designed with \callmeryd{} to engage young adults participating in the leadership academy and provide novel and creative methods for self-expression. As we iteratively refined what materials and processes supported putting data physicalization workshops into a metaphorical box, we also noticed what was perpetually difficult to put into a box.
We posit that these reflections were made more apparent to us as workshop observers, as opposed to workshop facilitators, precisely because \wib{w} provided space to observe workshops from a perspective often unafforded or opaque to visualization researchers. Our reflections broadly cover two topics. The first covers the difficulty of putting a data-to-visualization pipeline --- which underpins all visualization processes --- into a box. Next, we discuss the complexities of data production and what it means for both workshops and socially-driven applied visualization research.

\vspace{-1em}
\subsection{Data Encoding}
While there are many truths to the statement \textit{visualizing data is difficult}, we specifically emphasize the challenges we came across as we tried to put data physicalization workshops into a box: \textit{data encoding} and \textit{data production}. In our design study, we worked closely with our collaborator to develop data physicalization activities that were meaningful both for his organization's purpose and also for the young women in \organization{}'s youth leadership academy. Before each workshop, we would spend at least one meeting brainstorming the questions we would ask and building a visualization workshop around. Surprisingly, this was very difficult. From our perspective, we had fuzzy ideas that were too abstract to operationalize. \callmeryd \hspace{0.025cm} offered examples of the types of things he and his colleagues were interested in learning about, but they were often too large in scope to break down into concrete and datafiable questions. Over the course of the pilot workshops, we repeatedly found it difficult to co-develop meaningful and interesting data encodings. Our suspicion that this was a difficult problem was confirmed when we spoke to another colleague at \organization{} who mentioned that she wants to use the workshops in other settings, but would want help to engineer the question and ways of collecting data about the topic. Datafying a problem in the world became a part of the workshop that we ultimately could not put in a box. 

It was not only datafication that was a challenging component of \wib{w}, but also the development of abstract visual representations. As a research team, we suggested and built out the visual encodings. This limits the reusability of the \wib{w} to use cases that match or are closely related to those that surround the original context and motivation of the workshop. We saw similar limitations with the educators at the \visc{} who, although they have continued to be the most creative in using the physicalization workshops, the structure of the workshop and questions asked mirror the original pilot workshops. 

These limitations trouble the idea that putting a workshop in a box is enough to ethically exit~\cite{meyer2020criteria} when data visualization processes are still inaccessible to many people. But this begs the question. One research team is not enough to support data physicalization activities. We need a concerted effort to communicate visualization science research to the public to increase the skills and knowledge of how to abstract and encode data into visual representations. On the way there, though, we suggest these pragmatic steps. The visualization community could offer a collection of data physicalization ideas, abstracted to support a variety of needs and use cases. Or, we could consider designing a series of LLM prompts that result in useful modifications for physicalization activities.

\vspace{-1em}
\subsection{Data Production}
Another challenge that we repeatedly faced was the tension between pre-collected, situated, relevant data that promised the potential for more interesting physicalizations but was inconsistently collected compared to data that was easily created during the workshop but perhaps limited in its representational capacity. In the last two pilot workshops, it became clear that getting individuals to collect data before the workshops often resulted in a wide range of collected data. Some people collected a lot of data, others little, while some none at all, despite building in scaffolds like silent alarms. During the workshops where participants created data in the moment, there were no issues, and instead offered the opportunity to discuss the constructedness of data. 

We wondered how we could encourage data collection in future instances that might require in-situ data. In the field of personal informatics, scholars have discussed that personal tracking is difficult. Rapp \& Cena~\cite{rapp2016personal} worked with individuals, similar to our workshop participants, who were generally positive about collecting data but had no specific reason other than the study. After conducting a diary study, the researchers reported that individuals found data tracking \textit{"cumbersome, annoying, unsuitable, complex, stressful"} and either forgot to or had little motivation to input. This holds true for automated data tracking with apps like smart watches, that are eventually disposed of and forgotten in a drawer~\cite{epstein2016reconsidering}. 

This challenge poses two critical research paths for visualization researchers seeking to conduct applied visualization research outside of labs and other formal data analytics settings. First, there is theoretical and empirical work to be done to understand the data collection practices of individuals who use visualizations \textit{casually} outside of work. How do different motivations, attitudes toward data and visualizations, and technical expertise impact the ways that individuals approach data collection for visualization? While there is much theoretical and empirical work surrounding personal informatics, we call for visualization-specific studies to understand the nuances as pertaining to visualization. Second, we question the power of data collection if the collection is difficult to begin with. At the core of data humanism~\cite{lupi2017data} --- which inspired the design of our workshops --- is the idea that small, personal data can be powerful, meaningful, and impactful. Like others who faced similar disappointment~\cite {desjardins2020parallels}, we began with a naïve optimism that overlooked the challenges and realities inherent in data collection. How can data be powerful if people don't want to collect it? Rather, we offer a different interpretation of this maxim. In our workshops, we found that the \textit{process} of creating and visualizing data together was often more important and insightful than the outputs themselves. As such, we encourage other researchers to use data and visualization activities for purposes \textit{other than} outputting a dataset and a visualization, and instead as a mediator for conversations.
\vspace{-3em}

\rev{\section{Limitations}}

\rev{Both the pilot workshops and validation case studies were conducted as data physicalization workshops in community-oriented settings, which opens up the question of the generalizability and transferability of \wib{w} to other contexts~\cite{tracy2010qualitativea}. We hold that our strategies for putting knowledge in a box are deeply influenced by the material nature and process unique to data physicalization. While we speculate that there would be overlapping considerations with other forms of visualization workshops, we imagine that these use cases would require other sets of explicit strategies. For instance, if a workshop were to utilize visualization software, then the \wib{w} would need to externalize the tacit and procedural knowledges that are necessary for working with digital technologies.} 

\rev{We also speculate about the extent our recommendations are specific to community-oriented settings. Although it is possible that \wib{w} may work in other more general community workshops, we hold that the specifics of the workshop --- the format of the data, the type of visualizations, the ultimate purpose, and goals of the community partner --- would require further expansion to our suggested strategies to be rendered useful.}

\section{Conclusion}
Working closely with a social innovation organization led us to develop and deploy data physicalization workshops that our collaborators could facilitate independently. To put the workshop in a box, we iteratively refined the set of material and procedural considerations. By trying to think inside the box, we had the space to observe everything outside it. 
Removing the visualization researcher from the facilitation of the workshop opens up new opportunities for others to use data and visualizations in novel ways, such as tools for self-expression, catalysts for reflections, or mediators of discussions. Critically, by putting workshops in a box, we challenge the perception that data visualizations are solely tools for professional and educational contexts and instead create opportunities that center their use in our everyday lives. 


\section{Acknowledgments}
This work is funded in part by the Wallenberg
AI, Autonomous Systems and Software Program (WASP) funded by
the Knut and Alice Wallenberg Foundation; Svenska Postkod Lotteriet; and by InspireLab at KTH. We would also like to thank the people and spaces who made the workshops possible, fun, and interesting, including the Visualiseringscenter C, Östergötlands Stadsmission, Maria-Paz Briones, Lisa Sundén, and Matilda Stafsted.

\bibliographystyle{eg-alpha-doi} 
\bibliography{main}

\end{document}